# The nature of loops in programming

(*Down the mountains of the Noetherlands*)

## Bertrand Meyer

Bertrand.Meyer@inf.ethz.ch

*Abstract*: In program semantics and verification, reasoning about loops is complicated by the need to produce two separate mathematical arguments: an invariant, for functional properties (ignoring termination); and a variant, for termination (ignoring functional properties). A single and simple definition is possible, removing this split. A loop is just the limit (a variant of the reflexive transitive closure) of a Noetherian (well-founded) relation. To prove the loop correct there is no need to devise an invariant and a variant; it suffices to identify the relation, yielding both partial correctness and termination. The present note develops the (small) theory and applies it to standard loop examples and proofs of their correctness.

## 1 THE GOAL: AVOIDING A PERSONALITY SPLIT

Loops are key to software, since they leverage computers' power to repeat program elements many times and very quickly. In software verification, the dominant way to prove the correctness of a loop relies on a technique introduced by Hoare [10], taking over from Floyd [7], which requires you simultaneously to pull *two* rabbits out of a hat and prove their properties separately:

- A **loop invariant**, a boolean expression which you must prove to be satisfied on initialization and preserved by every iteration, showing that *if the loop terminates* the final state will still satisfy that invariant. This part of the proof addresses "partial correctness", meaning correctness *conditional* on termination.
- A **loop variant**, a non-negative integer expression which you must prove to decrease at every iteration. This part of the proof addresses termination, irrespective of what the loop actually does.

For example, the well-known binary search algorithm, which finds out whether a number $x$ appears in an array $t\ [1..n]$ of numbfaers known to be in increasing order, works by cutting the interval $1..n$ repeatedly in half and comparing $x$ to the value at the mid-point, going right if $x$ is bigger and left if $x$ is smaller, until it either finds $x$ as one of these midpoints or is unable to cut the interval further. Its invariant is the property that "$x$ appears in the array if and only if it appears in the current interval". Its variant is the size of the interval at each iteration. Figure 1 illustrates the search – respectively successful and unsuccessful – for $x = 61$ or $x = 62$ in the array shown.

Applying this division requires developing a split personality, "Dr. Termination and Mr. Correctness". The two jobs are as different as proctoring students during an exam and grading their answers. The split is probably part of why the concepts can be hard to teach (as illustrated by a naïve online discussion between programmers on this topic [21]). It is well entrenched in software verification, as part of the more general habit of dealing separately with "liveness" and "safety" properties going back to work of Lamport and Schneider [1][19].

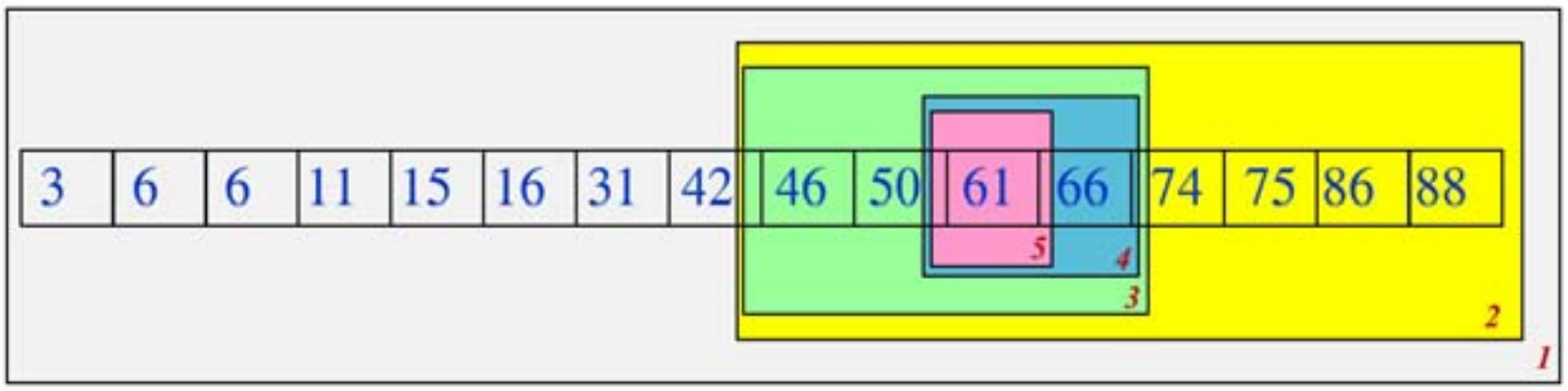

**Figure 1: Successive intervals in binary search**

Could we remove the split? In his guarded-commands semantics [5] [6], Dijkstra famously tried. He expressed the semantics of a loop through a *weakest precondition* $L$ **wp** $Q$ for a loop $L$ and postcondition $Q$, given by an existential formula ($\exists$ i: $\mathbb{N}$ | $H_i$ where the sequence $H$ is defined inductively by $H_0 = Q \wedge \neg U$, defining $U$ as the "or" of the guards, and $H_{i+1} = (I \textbf{ wp } Hi) \vee H_0$, where $I$ is the conditional with the same components as the loop, whose "**wp**" is also defined by a non-trivial formula!). Everyone was in awe at the formula's cleverness, and everyone made sure to forget it quickly as it is useless for practical verification. Tools typically rely on Hoare's separate "invariant + variant" technique. This note revives the challenge. It proposes a solution for the goal that eluded Dijkstra: unifying the semantics and verification of loops into a single rule.

With his signature modesty, Dijkstra qualified his solution as "*simple and elegant*" [6]. Authors do not usually bestow such praise on their own solution but leave it to others. We will therefore not follow his lead but just note that the interpretation of loops in the present work relies on a single concept, explained in the next sections: a loop is a procedure for finding the minimum of a Noetherian order by following a chain in a sub-relation.

As a preview, its application to binary search, or any search method of the same general style, seeks a minimum of the Noetherian "superset" relation ("$\supset$") on the set of subintervals of $1..n$ such that the corresponding subarray contains $x$ if an only if the full array does. A procedure (such as binary search) that starts from $1..n$ and iteratively replaces the current interval by a smaller one is guaranteed to yield a minimum of the relation. That minimum answers the search question ($x$ appears in the array if and only if the interval's size is zero). Using Noetherian relations merges the variant and invariant into a single argument.

*Novelty*: The connection between loops and well-founded sets is well-known, starting with an early article by Morris [16]; it is common lore in verification that loop variants could be chosen from any well-founded set rather than just natural integers. The classic article on termination [3] refers repeatedly to the concept. (Turing, in the first ever paper on program verification [23], used ordinals.) Abstract interpretation also uses well-founded relations [4]. Such discussions, however, typically refer to well-foundedness solely to prove termination. The originality of the present work is to show that Noetherian relations and orders suffice to capture *all* properties of loops, removing the separation between correctness and termination.

## 2 About loops

A program or program step may be defined as the combination of a set, the "precondition", and a relation between input and output states, the "postcondition"; see [15] for the full exposition. For simplicity and focus the present discussion ignores the precondition. An instruction such as a loop or loop body is then simply a relation between states.

In a loop **from** *i* **until** *C* **loop** *b* **end**, *C* (a subset of the set of states) is the "exit condition"; its complement (negation) *C'* is the "continuation condition". Mathematically, the loop is simply

$$i \; ; (C' : b)^*$$

where ";" denotes composition of relations, $C: r$ the restriction of a relation $r$ to a subset $C$ of its domain, and "*" reflexive transitive closure. $i$ is the loop's initialization and $b$ its body. (An even shorter formula will appear in section 3.4.)

(*Reminder and notations*: A relation $r$ on a set $X$ is a set of pairs $[a, a']$ of elements of $X$. Its inverse $r^{-1}$ is the corresponding set of $[a, a']$ pairs. Its domain $\underline{r}$ is the set of elements $a$, and its range $\overline{r}$ the set of elements $a'$, such that at least one such pair exists. Its transitive closure $r^+$ is the union $\bigcup_{0 < n} r^n$ of the powers of $r$, themselves defined by $r^1 = r$ and $r^{n+1} = r \; ; r^n$. The *reflexive* transitive closure $r^*$ is $r^+$ extended with $r^0$, defined as the identity relation $Id\,[X]$, the set of pairs $[a, a]$ for all $a$ in $X$. The image by $r$ of an element $a$, written $r\,(a)$, is the set of $a'$ such that $[a, a'] \in r$.)

Like any program or program step, the loop serves to fulfill a certain specification $s$; more precisely, it refines or specializes $s$, meaning simply in the present context that $r \subseteq s$ (subsetting between sets of pairs); [15] provides the details.

The loops of concern for this discussion are those which always terminate. (In operating systems and other applications some loops are repeating forever, at least in principle, but loops as covered here are an algorithmic mechanism to obtain a result.) This condition requires that the relation $C' : b$, as used in the formula defining the loop, be **finitary** as defined next.

When using relations to model components of loops and other programs, we note that for *deterministic* programs $r$ is a function (meaning that $r\,(a)$ always has 0 or 1 element).

## 3 Noetherian relations

The discussion will use a relation called ">", with $a > a'$ expressing that the pair $[a, a']$ belongs to the relation. ">" and "<" without the red color denote the usual order relations on numbers.

### 3.1 Chains

A **chain** is a non-empty sequence $s$ of elements, finite or infinite, such that $s_i > s_{i+1}$ for all $i$.

A relation is **Noetherian** (synonymously, *well-founded*) if it has **no infinite chain**. (Equivalently: any chain is finite.) Fig. 2 illustrates the idea: all chains hit a "wall".

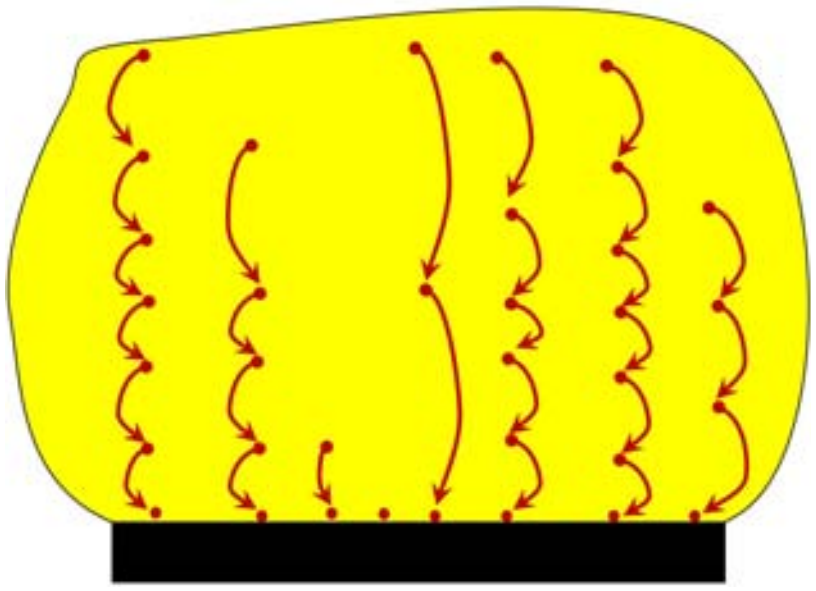

**Figure 2: With a Noetherian relation, all chains are finite**

A **minimum** of a Noetherian relation $r$ is an element $a$ that is not in its domain $\underline{r}$ (equivalently, ($r\,(a) = \varnothing$).

## 3.2 Properties

(*Reminder*: A relation is acyclic if it has no cycles; formally, $r^+ \cap Id\,[X] = \varnothing$, since a cycle would be obtained by applying $r$ one or more times and getting back to the same element. A relation is an order relation if it is irreflexive and transitive; as a consequence it is also asymmetric since $a > a'$ and $a' > a$ would imply by transitivity that $a > a$, defeating irreflexivity. By the same argument an order relation is acyclic. We are talking here of *strict* order relations, using "$>$"; a non-strict order, "$\geq$", can be defined from its strict counterpart, as $a > a' \vee a = a'$, or on its own as reflexive, antisymmetric and transitive. In both cases, strict and not strict, the order relations considered here are *possibly partial*: given two different elements, it may be the case that $a > a'$ or $a < a'$ but also that neither of these holds.)

The Noetherian property is usually applied to *order* relations. The above definition is more general and is the one we need for a loop body, which is generally not an order (only the loop as a whole, as a transitive closure, is). A Noetherian relation does have some of the properties of a (strict) order relation:

- It is acyclic: if there were a cycle $a > b > \ldots > n > a$, we could follow it forever, in an infinite chain. This proof does not require transitivity (as used in its counterpart for order relations). Note that, the other way around, not all acyclic relations are Noetherian. (See "$>$" in $\mathbb{Z}$.)
- As a consequence, it is irreflexive (no cycles of length 1) and asymmetric (none of length 2).
- Any of its positive powers and its (non-reflexive) transitive closure is Noetherian. (Proof: a chain in $r^i$ (for $i > 0$) or $r^+$ is a chain in $r$.) In the application to loops, this property is important since the transitive closure of an acyclic relation is an order relation (it is irreflexive like all acyclic relation, and transitive by construction).
- The composition of two Noetherian relations is Noetherian.

## 3.3 Limits

We will say that a relation is **finitary** (short for "has finitary closure") if $r^*(a)$ is obtained, for any element $a$ of $X$, by applying a finite union $\cup_{0 \le i \le M} r^i$ for some natural integer $M$. For example ">" is finitary on $\mathbb{N}$ but not on $\mathbb{Z}$ (where, starting from an integer, you can forever get smaller elements). This $M$ may depend on $a$; it is called the "height" (or "rank") of $a$ for $r$ and will be written it $M_r(a)$, with the subscript $r$ ommitted if clear from the context. To obtain the reflexive transitive closure for $a$ it $M(a)$ iterations of $r$ suffice.

Any Noetherian relation is finitary. (Proof: $M(a)$ is the maximum length of a chain starting at $a$.) As a consequence, for Noetherian $r$ and any $a$, $r^{M(a)}(a) \cap \underline{r} = \varnothing$. (Proof: this were not the case, $r^{M(a)+1}$ would be non-empty, but since the relation is acyclic its elements would not belong to previous powers of $r$, defeating the finitary hypothesis.) In other words, elements of $r^{M(a)}(a)$ are **minima** of both $r$ and $r^*$.

The relation that yields the elements of $r^{M(a)}(a)$ for every $a$ will be called the **limit** of $r$ and written $r^\bullet$. Figure illustrates $r^\bullet$ for the relation of Figure 3.

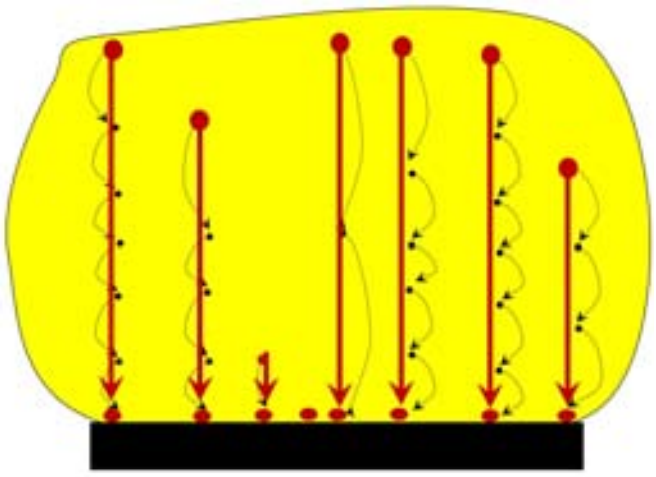

**Figure 3: Limit of the relation of Fig. 2**

In general, $r^\bullet$ is not transitive. Some immediate properties are: $r^\bullet \subseteq r^*$; $(r^*)^\bullet = r^\bullet$; any chain of $r^\bullet$ has length 0 or 1. The *limit subset theorem* states that if $r$ is a *seed* of $s$, defined as $r \subseteq s$ and $\underline{r} = \underline{s}$ (the two relations have the same domain), then $r^\bullet = s^\bullet$. (Proof: a minimum — end of a chain — of $r$ is also a minimum of $s$ since it is not in the domain of $r$ and hence not in the domain of $s$.) Typically $r$ will be a much smaller relation than $s$; in applications of this rule $s$ is an order but $r$ is not transitive. For example, on $\mathbb{N}$, is ">" and $s$ is *predecessor* (the relation with all $[n+1, n]$ pairs).

For a deterministic program, as noted, $r^\bullet$ yields *the* minimum associated with $a$ (the last element of the — in this case unique — chain of $r$ starting with $a$).

## 3.4 What you always wanted to know about loops (but did not know how to ask)

The preceding concepts provide a proper background for understanding and building loops.

In the spirit of treating control structures as problem-solving technique [6] [9] [11], a loop $l$ is a minimization procedure. It seeks a minimum of a Noetherian order ">" on a set $X$; more precisely, for any given $a$ in $X$, a minimum element $m$ such that $a \ge m$. It is defined by four ingredients:

- A non-empty subset $X$ of the set of states.
- A Noetherian order "$>$".
- An initialization $i$ such that $\bar{i} \subseteq X$ (so that $i$ yields an initial element in $X$).
- A body $b$. which is a seed of "$>$".

A loop $l$ is

$i \,;\, b^{\bullet}$

which yields minima of "$>$" (from the above property that if $r^{\bullet} = s^{\bullet}$ if $r$ is a seed of $s$). Viewed operationally as a program, the loop:

- Starts by applying $i$ to the initial state.
- Executes $b$ as long as possible, that is to say as long as the state is in the domain of $b$, which is also the domain of $l$.
- When it stops, has found a minimum of $l$.

Each of the examples of section 5 defines a loop (and hence its proof of correctness) by listing the four components above. It also indicates two properties that do not need to be provided to define a loop since they follow from those four:

- The postcondition of the loop (the property that it achieves), which is the characterization of minima of the "$>$" relation.
- The exit condition $C$, which is $X \cap \underline{b} = \varnothing$.

The exit condition, or its complement C', appear in programming-language forms of the loop:

**from** $i$ **until** $C$ **loop** $b$ **end**

or

$i$ ; **while** $C'$ **loop** $b$ **end**

### 3.5 Mathematical variants

We may note for completeness that the mathematical framework defined above is one of four possible variants. Instead of a "greater than" relation it is possible to use "less than"; then a Noetherian relation is one for which there are no ascending chains (rather than no infinite descending chains as used here). Orthogonally, it is possible to use *non-strict* variants of order relations, such as "$\geq$" (greater than or equal) instead of "$>$" and similarly in the other direction. Then the Noetherian condition is that there are no stationary (rather than no infinite) chains, where a chain is said to be stationary if beyond a certain index all elements are equal.

It is possible to develop the theory with any one of the four variants. We will stick to the strict "greater than" version "$>$" and the corresponding no-infinite-descending-chain condition.

## 4 Examples of Noetherian relations

The following are examples of Noetherian relations, which will provide with a rich starting pool of techniques for finding loop solutions.

The specification for each defines under what condition $a > a'$ holds. To avoid any confusion, note the following conventions: each relation is defined in the direction "greater than"; the condition specifying when $a > a'$ holds always lists $a$ first and $a'$ second; and the name of the relation is chosen to indicate what $a$ is in relation to $a'$. For example, in a tree, the relation that holds when node $a$ is the parent of node $a'$ is called "[CHILD]" because when applied to $a$ it yields the set of children of $a$, to which $a'$ belongs. Similarly, the relation that holds between integers when $a = a' - 1$ is called [SUCCESSOR].

"Interval" is short for "finite integer interval"; an interval is written $a..b$ where $a \leq b + 1$, and $a = b + 1$ indicates an empty interval.

The first few entries are general mechanisms for obtaining new Noetherian relations from known ones (including two already mentioned, here given names); subsequent relations are more specific. Relations with an <u>underlined</u> name are Noetherian orders.

[COMPOSE]

Composition of two Noetherian relations on the same set. Not necessarily an order even if both are.

[<u>CLOSURE</u>]

Transitive closure of a Noetherian relation.

[SUBREL]

Subset of a Noetherian relation. (Proof: if $r \subseteq s$, then any chain for $r$ is a chain for $s$.) Not necessarily an order even if the original is.

[RESTRICT]

$C: r$ where $r$ is Noetherian. (Special case of [SUBREL].) Not necessarily an order even if $r$ is. Example: "$>$" on some proper subset of $\mathbb{N}$. Proofs of loop correctness in section 5 often implicitly rely on [RESTRICT] by using one of the other relations and restricting it to a specific subset.

[INDUCED]

$f(a) > f(a')$ where $f$ is a function from $X$ to $Y$ and "$>$" is Noetherian on $Y$. (Proof: the image by $f$ of any chain on $X$ is a chain in $Y$.) An order if the original is. The traditional use of a variant is an extreme case of this rule since it uses a function with a natural-integer values, using the Noetherian order [INTGREATER] below, to establish termination.

[PROJECTION]

If $X$ is a cartesian product $A \times B \times \ldots \times E \times \ldots$, and $r$ is a Noetherian relation on one of the components, say $E$, the relation defined by $r(e) > r(e')$ on the $E$ components. (Proof: this case falls under [INDUCED] where the function $f$ is the projection on the $E$ dimension.) An order if $r$ is. This rule is particularly important since states are typically made of many components, and it allows us to find a Noetherian relation on just one of them.

[INVERSE]
: Inverse of a Noetherian relation on a finite set. (Proof: the reverse of a chain is a chain in the original.) An order if the original is.

[ACYCLIC]
: Both $a$ and $a'$ are nodes in a finite acyclic graph and there is an edge from $a$ to $a'$. (As a consequence, any finite order relation (since order relations are acyclic). The underlying set can be infinite as long as the relation is finite.

[ACYCLIC']
: Both $a$ and $a'$ are nodes in a finite acyclic graph and there is an edge from $a$ to $a'$. (Consequence of [ACYCLIC] and [INVERSE].)

[PARENT]
: $a$ is the parent of $a'$ ($a'$ is a child of $a$) in a tree (rooted acyclic graph, not necessarily finite) or forest (set, not necessarily finite, of trees).

[ANCESTOR]
: Transitive closure of [PARENT].

[CHILD]
: $a$ is a child of $a'$ ($a'$ is the parent of $a$) in a finite-depth tree or forest. (Nodes may actually have an infinite number of children but the depth must be finite.) Inverse of [PARENT].

[DESCENDANT]
: Over a finite-depth tree or forest: transitive closure of [CHILD], inverse of [ANCESTOR].

[SUPSET]
: The strict superset relation "$\supset$"on a finite set. (Special case of [ACYCLIC'].)

[SUBSET]
: The strict subset relation "$\subset$"on a finite set. (Follows from [SUPSET] and [INVERSE].)

[SUCCESSOR]
: $a = a' + 1$ on $\mathbb{N}$ (natural integers).

[INTGREATER]
: "$>$" on $\mathbb{N}$. (Transitive closure of [SUCCESSOR].)

[PREDECESSOR]
: $a = a' - 1$ on a subset of $\mathbb{Z}$ having an upper bound.

[INTLESSER]
: "$<$" on a finite subset of $\mathbb{Z}$. (Transitive closure of [PREDECESSOR], inverse of [INTGREATER].)

[INTDIFF]
: On pairs of integers: $|m - n| > |m' - n'|$.

[INTSUM]
: On pairs of natural integers: $m + n > m' + n'$.

[MAXINT]
: On pairs of natural integers: $max\ (m, n) > max\ (m', n')$.

[MININT]

On pairs of natural integers: $min\ (m, n) > min\ (m', n')$.

[SUPINTERVAL]

On intervals: $a < a' \vee b > b'$. (Special case of [SUPSET], also of [INTDIFF].)

[SUBINTERVAL]

On subintervals of a given $m..n$: $a > a' \vee b < b'$. (Special case of [SUBSET], inverse of [SUPINTERVAL].)

[INTERVAL]

On intervals: [INTDIFF] applied to the intervals' bounds: $|a - b| > |a' - b'|$. ([SUPINTERVAL] is a special case.)

[INTERVAL']

On subintervals of given $m..n$: inverse of [INTERVAL]: $|a - b| < |a' - b'|$.

[INTERVALSUPSET]

On a set of intervals: [SUPSET]. (This relation and the next not to be confused with the preceding ones which apply to intervals, not sets of intervals.)

[INTERVALSUBSET]

On a set of subintervals of given $m..n$: [SUBSET].

[INTERVALMAX]

On a finite set of intervals, their maximum length.

# 5 LOOP EXAMPLES AND THEIR CORRECTNESS PROOFS

We apply the previous concepts to a number of important and typical examples of loops. In examples using an array $t\ t\ [1..n]$: the notation $t\ [i..j]$ represents the subarray having the given bounds; $x < t\ [i..j]$ means that $x$ is less than all the elements of that subarray; similarly, $t\ [i..j] < t\ [k..l]$. that all elements of the first slice are less than all elements of the second slice. In such inequalities, $t$ is an abbreviation for the entire slice $t\ [1..n]$.

## 5.1 Euclid's gcd algorithm

Using the version with subtraction (division would also work), input is $[a, b]$ both positive:

| Set | Order | Initialization | Body |
|---|---|---|---|
| Pairs $[m, n]$ of positive integers such that $gcd\ (m, n) = gcd\ (a, b)$ | [MAXINT] (or [INTSUM]) | $m := a$<br>$n := b$ | **if** $m > n$<br>**then** $m := m - n$<br>**else** $n := n - m$ **end** |

Exit condition: $m = n$. Postcondition: $m = n = gcd\ (a, b)$.

## 5.2 Searching in an array

Sequential search traverses an array $t\ [1..n]$ from left to right until it either finds a sought elements $x$ or reaches the end without having found it (Fig. 4).

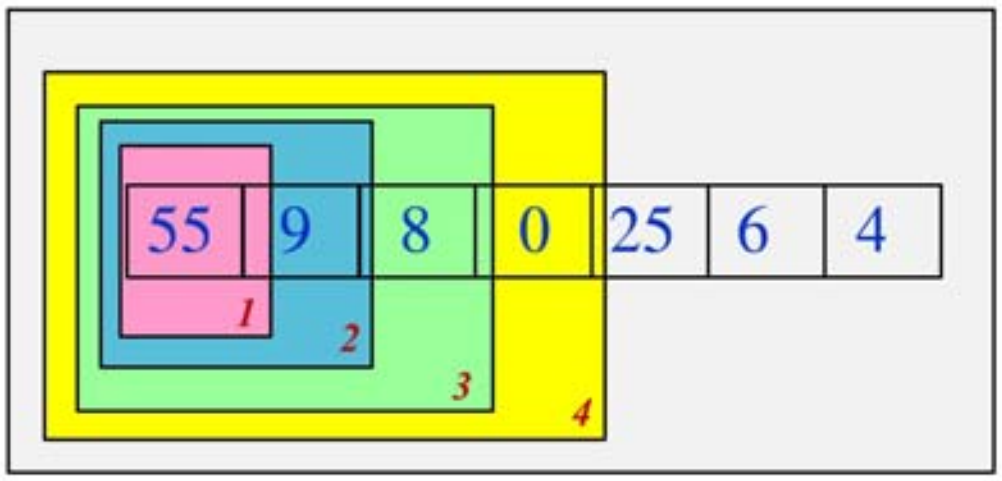

**Figure 4: Intervals in sequential search**

| Set | Order | Initialization | Body |
|---|---|---|---|
| Intervals $1..i$ for $0 \le i \le n$ such that $x$ does not appear in $t\ [1..i]$ | [SUBINTERVAL] | $i := 0$ | $i := i + 1$ |

Exit condition: $i = n$ **or else** $t\ [i + 1] = x$. Postcondition: $(x \in t) = (i \ne n)$.

## 5.3 General search

Not all search algorithms use, as sequential search does, an increasing interval. A more general solution, which applies to many search algorithms including binary search (Fig. 1), uses a decreasing interval:

| Set | Order | Initialization | Body |
|---|---|---|---|
| Subintervals $I$ of $1..n$ such that $(x \in t) = (x \in t\ [I])$ | [SUPINTERVAL] | $I := 1..n$ | $I := J$ |

$J$ is a (strict) subinterval of $I$. Exit condition: $I = \varnothing \lor (x \in t\ [I])$. Postcondition: $(x \in t) = (I \ne \varnothing)$.

An alternative model and proof for such search algorithms use a *set* (increasing) of intervals:

| Set | Order | Initialization | Body |
|---|---|---|---|
| Sets $IS$ of subintervals of $1..n$ such that $x \notin t\ [I]$ for all members $I$ of $IS$ | [INTERVALSUBSET] | $IS := \varnothing$ | $IS := IS \cup \{J\}$ |

$J$is a subinterval not in $IS$. Exit condition: $(\bigcup IS = 1..n) \lor (x \in J)$. Postcondition: $(x \in t) = (\bigcup IS \ne 1..n)$.

## 5.4 Partitioning for Quicksort

Quicksort's Partition procedure moves to the left elements smaller than a pivot, and to the right elements that are greater than it (Fig. 5).

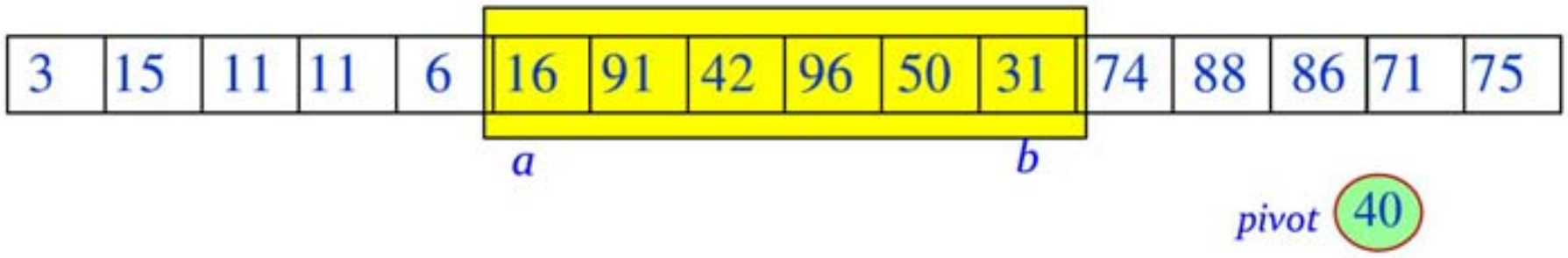

**Figure 5: Array being partitioned for Quicksort**

| Set | Order | Initialization | Body |
|---|---|---|---|
| Subintervals $a..b$ of $1..n$ such that for an array that is a permutation of the original one, $t\,[1..a{-}1] \le pivot$ and $t\,[b{+}1..n] \ge pivot$ | [SUBINTERVAL] | $1..n$ | **if** $t\,[a] \le pivot$ **then**<br>$a := a + 1$<br>**elseif** $t\,[b] \ge pivot$ **then**<br>$b := b - 1$<br>**else**<br>"$t\,[a] \leftrightarrow t\,[b]$"<br>$a := a + 1$<br>$b := b - 1$<br>**end** |

Exit condition: $a = b + 1$. Postcondition: $t$ is a permutation of the original and $t\,[1..a{-}1] \le t\,[a..n]$.

## 5.5 Lampsort (Quicksort)

Popular opinion considers Quicksort itself to be a recursive algorithm but Leslie Lamport convincingly argued that there is nothing inherently recursive in it. The "Lampsort" version, explained in [13], keeps a set of non-empty adjacent subintervals covering $1..n$., such that the array remains a permutation of the original one and $t\,[I] \le t\,[I']$ for adjacent intervals (Fig. 6). At each step it picks one of them and partitions it into two smaller intervals using the "partition" procedure covered above.

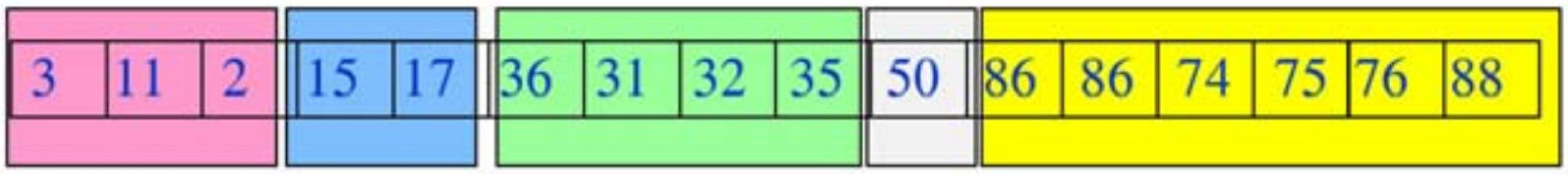

**Figure 6: Lampsort intermediate state**

This algorithm cries for Noetherian modeling, with a trivial proof:

| Set | Order | Initialization | Body |
|---|---|---|---|
| Sets of adjacent intervals $I$ covering $1..n$, such that $t$ is a permutation of the original and $t[I] \leq t[I']$ for every adjacent $I$ and $I'$. | [INTERVALMAX] | $\{1..n\}$ | Partition one $I$ of length 2 or more |

Exit condition: All $I$ have length 0 or 1. Postcondition: $t$ is a permutation of the original and sorted.

### 5.6 Other examples

Further versions of this note will include examples involving linked data structures and further delicate cases from [3] and [8].

## 6 Discussion

The application of the presented approach to practical program verification assumes that we have a way to associate with a loop the Noetherian relation that it implements.

The general principle is the idea of model specifications discussed in [20]and [18] and implemented in the Autoproof program verification environment for Eiffel [22] [2] through the Mathematical Model Library (MML). The idea is simply to associate with any program object a mathematical object, obtained through MML, typically an integer (in simple cases), a set, a sequence, relation, a function or some combination of such elements. Then the effects of program operations can be specified in terms of the effects on the model's mathematical objects. For example, if a stack has at its model a sequence with the top as first element, then the postcondition for the "push" operation states that the new model sequence is equal to the old one with the new element prepended.

Applying the presented approach in this framework consists of defining the Noetherian relation as a model property associated with a loop. A proof consists principally of showing that the body is a Noetherian subset of that relation. Much of the work can be done once and for all, independently of specific programs, by adding to MML a library of widely useful Noetherian orders and other Noetherian relations, starting with those of section 4 above, and a library of their preproved formal properties, in particular which of these relations are subsets and seeds of others.

The approach subsumes. abstracts and unifies the traditional invariant + variant method. It has not been tried on any significant scale but may be promising.

# BIBLIOGRAPHY

[1] Bowen Alpern and Fred B. Schneider: *Defining Liveness*, in *Information Processing Letters*, Vol. 21, No. 4, pages 181–185, October 1985.

[2] AutoProof online proof environment, available at autoproof.constructor.org.

[3] Byron Cook, Andreas Podelski and Andrey Rybalchenko: *Proving Program Termination*, in *Communications of the ACM*, Vol. 54, No. 5, pages 88–98, May 2011, available at doi.org/10.1145/1941487.1941509.

[4] Patrick Cousot: *Principles of Abstract Interpretation*, MIT Press, 2021.

[5] Edsger W. Dijkstra: *Guarded commands, nondeterminacy and formal derivation of programs*, in *Communications of the ACM*, Vol. 18, No. 8, pages 453–457, 1975.

[6] Edsger W. Dijkstra: *A Discipline of Programming*, Prentice Hall, 1976.

[7] Robert W. Floyd: *Assigning meanings to programs*, in *Proceedings of the Symposium on Applied Mathematics*, Vol. 19, pages 19–32, American Mathematical Society, 1967.

[8] Carlo A. Furia, Bertrand Meyer, and Sergey Velder: *Loop invariants*: *Analysis*, *classification*, *and examples*, in ACM *Computing Surveys*, Vol. 46, No. 3, pages 1-51, 2014.

[9] David Gries: *The Science of Programming*, Springer, 1981.

[10] C.A.R. Hoare: *An axiomatic basis for computer programming*, in *Communications of the ACM*, Vol. 12, No. 10, pages 576–580, 1969.

[11] Bertrand Meyer: *A Basis for the Constructive Approach to Programming*, in *Information Processing* 80 (Proceedings of the IFIP World Computer Congress), Tokyo, October 6-9, 1980, ed. S. H. Lavington, North-Holland, pages 293–298, 1980.

[12] Bertrand Meyer: *Object-Oriented Software Construction*, *2nd edition*, Prentice Hall, 1997.

[13] Bertrand Meyer: *Lampsort*, blog article, 7 December 2014, available at bertrandmeyer.com/2014/12/07/lampsort.

[14] Bertrand Meyer: *Getting a Program Right*, *in Nine Episodes*, blog article, 26 March 2020, available at bertrandmeyer.com/2020/03/26/getting-program-right-nine-episodes/.

[15] Bertrand Meyer and Reto Weber: *Programming Really Is Simple Mathematics*, to appear in *Yuri Gurevich Festschrift*, ed. Guillermo Badia, Springer, June 2025, also arXiv preprint, 24 February 2025, available at arxiv.org/abs/2502.17149.

[16] James H. Morris: *Inferencing on well-founded sets*, in *Proceedings of the ACM Annual Conference*, pp. 465–473, 1970.

[17] F. L. Morris and C. B. Jones: *An Early Program Proof by Alan Turing*, in *Annals of the History of Computing*, Vol. 6, No. 2, pages 139–143, April 1984.

[18] Nadia Polikarpova, Carlo A. Furia, Yu Pei, Yi Wei and Bertrand Meyer: *What Good Are Strong Specifications?*, in 35*th International Conference on Software Engineering* (ICSE), pages 257–266, 2013.

[19] Fred B. Schneider and Leslie Lamport: A*nother Position Paper on* "*Fairness*", in *ACM SIGSOFT Software Engineering Notes*, Vol. 13, No. 3, pages 18–19, July 1988.

[20] Bernd Schoeller, Tobias Widmer, and Bertrand Meyer: *Making Specifications Complete Through Models*, in *Architecting Systems with Trustworthy Components*, eds. Ralf Reussner, Judith Stafford, and Clemens Szyperski, Lecture Notes in Computer Science, Vol. 3938, pages 48–70, Springer, 2006.

[21] *Stack Exchange* discussion on the theme "Finding a good loop invariant for a powering procedure", started 22 March 2016, available at cs.stackexchange.com/questions/54790/finding-a-good-loop-invariant-for-a-powering-procedure.

[22] Julian Tschannen, Carlo A. Furia, Martin Nordio, and Nadia Polikarpova: *AutoProof: Auto-active Functional Verification of Object-Oriented Programs*, in *Proceedings of the 21st International Conference on Tools and Algorithms for the Construction and Analysis of Systems* (TACAS), Vol. 9035, pages 379–398, Springer, 2015.

[23] Alan Turing: *Checking a Large Routine*, in *Report of a Conference on High Speed Automatic Calculating Machines*, University Mathematical Laboratory, Cambridge, pages 67–69, 1949. (Included, transcription-corrected and easier to read in [17], with an analysis.)

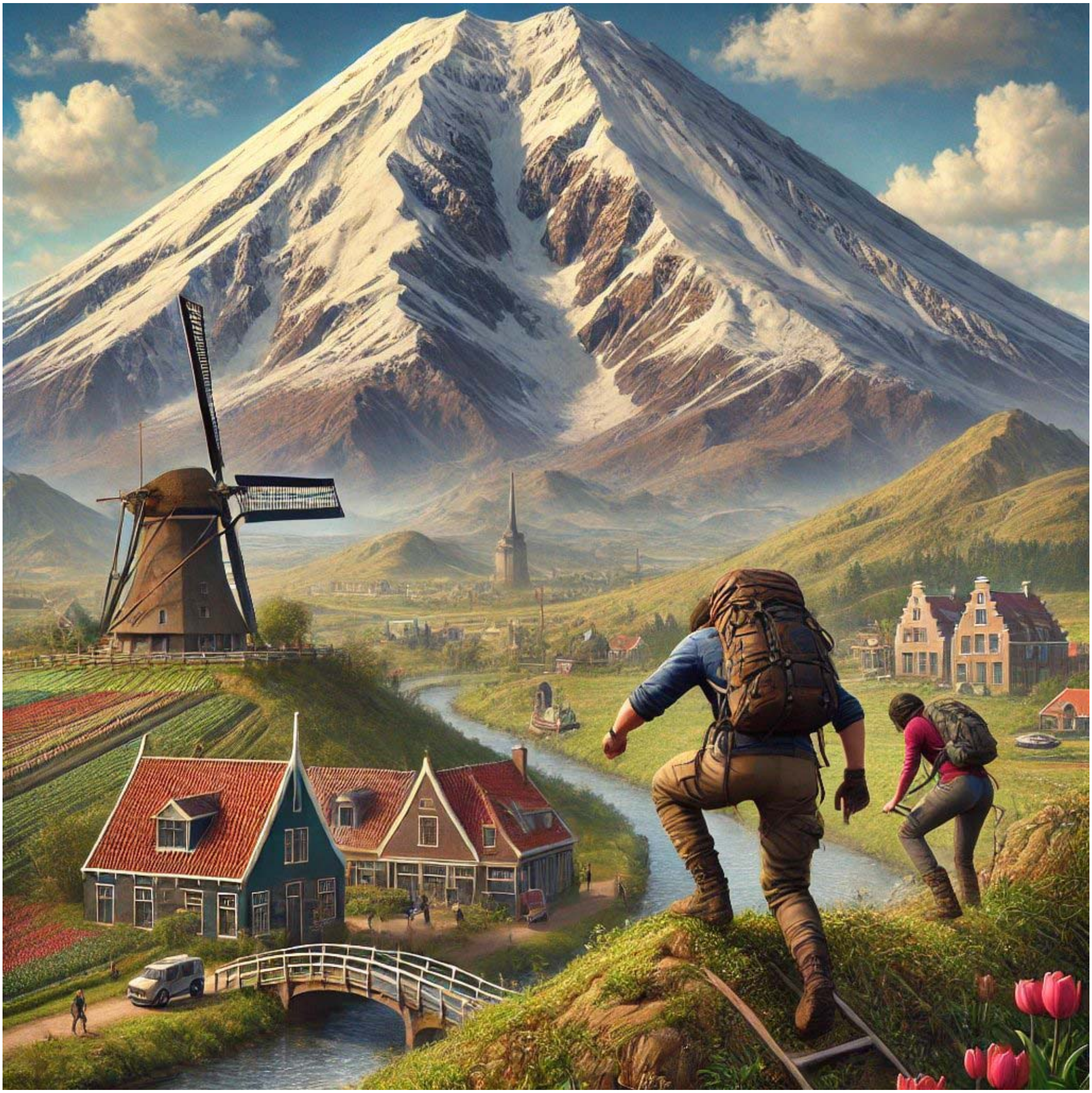